# Simulation of metallic nanostructures for emission of THz radiation using the lateral photo-Dember effect

Duncan McBryde, Mark E. Barnes, Geoff J. Daniell, Aaron L. Chung, Zakaria Mihoubi, Adrian H. Quarterman, Keith G. Wilcox, Anne C. Tropper, Vasilis Apostolopoulos

School of Physics and Astronomy, University of Southampton, Southampton, SO17 1BJ

*Abstract*—A 2D simulation for the lateral photo-Dember effect is used to calculate the THz emission of metallic nanostructures due to ultrafast diffusion of carriers in order to realize a series of THz emitters.

## I. Introduction And Background

THE photo-Dember effect can be used to create THz radiation by the difference of mobilities of holes and electrons [1]. Under photo-excitation in a semiconductor the different rates of diffusion of carriers create a fast changing dipole that gives rise to a THz wave. Under normal illumination with an ultrafast laser pulse the direction of the resulting dipole will be collinear to the direction of the laser, thus the THz radiation will be emitted perpendicularly to the optical excitation. Recently, Klatt et. al. [2] improved the performance of the photo-Dember emitter by demonstrating the lateral photo-Dember effect. It was demonstrated that by sharply masking the illuminated area, using a metal, the carriers diffuse laterally, causing the dipole to be created perpendicular to the surface normal and creating a terahertz pulse parallel to the direction of the laser pulse. It was also demonstrated that by using a series of gold ramps the power from multiple photo-Dember emissions could be combined resulting in enhanced output power by creating a net diffusion current in one direction. To create a net diffusion current in one direction there must be a break in the symmetry of the contrast gradient of the carriers. In [2] this was achieved by using gold ramps to vary the initial density of the carriers. We have developed a model of the lateral photo-Dember effect for the purpose of testing various masking structures. Our motivation is to create a repeating structure of constant thickness that can be fabricated by conventional lithography that can be scaled over a large area, so that many diffusion currents combine to give enhanced terahertz emission.

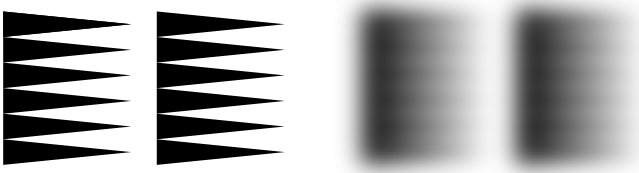

Fig. 1: The figure on the left shows a representation of the proposed structure. At low pitch spacing, electrons diffuse to form areas of high and low contrast.

## II. Theory

The model simulates the dynamics of electron and hole densities in the valence band of a semiconductor after photo-excitation by an infinitely short laser pulse. The holes are assumed to be static, as the ratio of the mobility of electrons to holes is high, such as 18 for GaAs, 71 for InAs for 104 in InSb [3, 4]. Electrons in the model are free to move, under the influence of electric fields and diffusion, and have a chance to recombine with holes if present. Under these assumptions the evolution for the hole density $n_h$ and the electron density $n_e$ with time $t$ can be written as,

$$\frac{\partial n_h}{\partial t} = -\frac{n_e n_h}{t_r}$$

$$\frac{\partial n_e}{\partial t} = \mu \vec{\nabla} \cdot (\vec{E} n_e) + D \nabla^2 n_e - \frac{n_e n_h}{t_r}.$$

Where $t_r$ is the average recombination time between holes and electrons, $\mu$ the mobility of electrons in the semiconductor, $\vec{E}$ the electric field and $D$ the diffusion constant of the electrons. The model is implemented as a two dimensional simulation with cyclic boundary conditions imposed on $n_h$, $n_h$ and $\vec{E}$, which corresponds to a unit cell in a repeating structure. In such a case the potential distribution $V$ is calculated from the charge distribution, with $\vec{E} = \vec{\nabla} V$. The potential $V$ is calculated by performing a Fourier transform on the charge distribution:

$$V(x, y) = \frac{1}{4\pi\epsilon} \mathcal{F}^{-1}\left(\frac{\mathcal{F}(e(n_h - n_e))}{k_1^2 + k_2^2}\right).$$

$\mathcal{F}$ and $\mathcal{F}^{-1}$ are the forwards and inverse Fourier transforms respectively, and $k_1$ and $k_2$ the dimensions in frequency space. By approximating the partial differential equations for $n_h$ and $n_e$ as finite differences while solving for $V$ the evolution of $n_e$ with time is obtained. In order to obtain qualitative results, the electric field of the terahertz emission is approximated as the second temporal derivative of the charge density.

$$E_{THz} \propto \frac{\partial^2 n_e}{\partial t^2}$$

## III. Results

The model was used to simulate the electric field emitted by a masking pattern of constant thickness. The patterns simulated were a repeating saw-tooth pattern as shown in figure 1. This is an attempt to create a similar carrier distribution to the wedges fabricated in [2] but without having the fabrication difficulties associated to a three dimensional structure. It was projected that if the pitch of the saw-tooth mask was small enough, electrons would diffuse in a short time to create the smooth carrier gradient, thus providing the needed asymmetry. The simulation produced net current resulting in electric field pulse profiles, as shown in figure 2. The majority of electrons remaining within the hole concentration recombine in a short amount of time, causing electrons to diffuse back to the original carrier concentration. By varying the ratio of pitch spacing to amplitude, the optimum amplitude of the saw-tooth mask was 800 nm with a pitch spacing of 200 nm as shown in figure 3. The simulated profile of the electric field of this structure has a bandwidth of 2.88 THz, measured as the FWHM of the amplitude spectrum, which is to be confirmed by experiment.

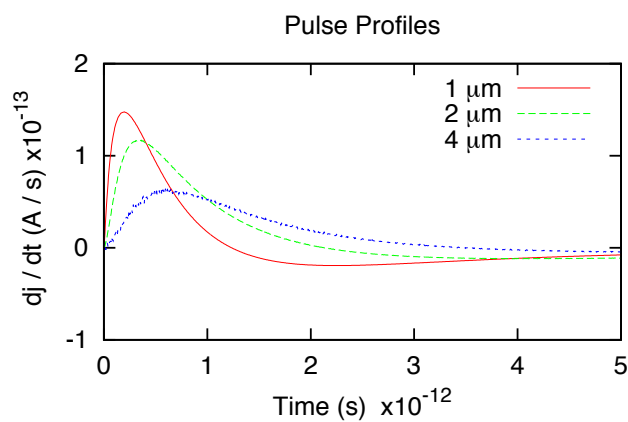

Figure 2: A graph showing the simulated electric field profile of saw-tooth emitters of various sizes with a 1:4 ratio of pitch to amplitude.

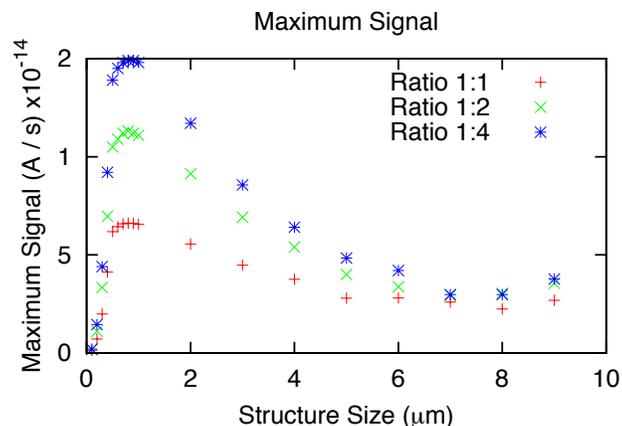

Figure 3: A graph of the maximum simulated electric field against structure size, with a comparison between different ratios of pitch spacing to amplitude. Peak emission occurs at a structure size of 800 nm with a pitch spacing of 200 nm.

However, by disabling the electric field, the net current in the simulation is zero, which indicates that net diffusion current is zero. Therefore, the THz pulses are a result of the asymmetric carrier profile, which creates an asymmetric electric restoring force for the electrons as they diffuse away from the area of the holes. The simulated diffusion of carriers from each side of the emitter, in fig. 1, is creating an equal current that cancels out. The current though produced from the electric field does not seem enough to create the THz emission observed in [2] as it is 4-5 orders less than the diffusion dipole created on each side of the emitter. This seems to indicate that there is another phenomenon that breaks the dipole symmetry and causes the observed THz pulse, which we have not taken into account.

## IV. Conclusion

The potential of the photo-Dember effect lies in the ability to scale the intensity of the terahertz radiation with size efficiently. The unit cells can be repeated over a large area producing enhanced emission and according to [2] it can produce comparable results to photo-conductive sources. It has the advantage that there is no need of electrical bias, so it is easier to fabricate and thus to integrate and will not have the lifetime problems associated with the electrical bias. The parameters for fabricating two-dimensional photo-Dember masks proposed here are possible using conventional electron-beam lithography or UV-based lithography. However, using our model we have demonstrated that the simulated two-dimensional masking pattern can generate terahertz radiation only due to the E-field and not due to diffusion currents. We are investigating the discrepancy between theory and experiment. Work is also currently underway to fabricate the saw-tooth based photo-Dember masks on GaAs and test for terahertz emission in comparison to our model.